\documentclass{emulateapj}
\usepackage{natbib}

\shorttitle{First detections of PO in star-forming regions}
\shortauthors{Rivilla et al.}

\usepackage{natbib}
\usepackage{graphicx}
\usepackage{amsmath}
\usepackage{epsfig}
\usepackage{appendix}

\begin{document}

\title{First detections of the key prebiotic molecule PO in star-forming regions}
%
\author{V. M. Rivilla\altaffilmark{1}, F. Fontani\altaffilmark{1}, M. T. Beltr\'an\altaffilmark{1}, A. Vasyunin\altaffilmark{2,3}, P. Caselli\altaffilmark{2}, J. Mart\'in-Pintado\altaffilmark{4}, and R. Cesaroni\altaffilmark{1}}      
\altaffiltext{1}{Osservatorio Astrofisico di Arcetri, Largo Enrico Fermi 5, I-50125, Firenze, Italia}
\altaffiltext{2}{Max-Planck Institute for Extraterrestrial Physics, Giessenbachstrasse 1, 85748, Garching, Germany}
\altaffiltext{3}{Ural Federal University, Ekaterinburg, Russia }
\altaffiltext{4}{Centro de Astrobiolog\'ia (CSIC-INTA), Ctra. de Torrej\'on a Ajalvir km 4, 28850, Torrej\'on de Ardoz, Madrid, Spain}
%
%

\begin{abstract}
Phosphorus is a crucial element in biochemistry, especially the P$-$O bond, which is key for the formation of the backbone of the deoxyribonucleic acid. So far, PO has only been detected towards the envelope of evolved stars, and never towards star-forming regions.
We report the first detection of PO towards two massive star-forming regions, W51 e1/e2 and W3(OH), using data from the IRAM 30m telescope. PN has also been detected towards the two regions. The abundance ratio PO/PN is 1.8 and 3 for W51 and W3(OH), respectively. Our chemical model indicates that the two molecules are chemically related and are formed via gas-phase ion-molecule and neutral-neutral reactions during the cold collapse. The molecules freeze out onto grains at the end of the collapse and desorb during the warm-up phase once the temperature reaches $\sim$35 K. Similar abundances of the two species are expected during a period of $\sim$5$\times$10$^{4}$ yr at the early stages of the warm-up phase, when the temperature is in the range 35$-$90 K. The observed molecular abundances of 10$^{-10}$ are predicted by the model if a relatively high initial abundance of 5$\times$10$^{-9}$ of depleted phosphorus is assumed.

\end{abstract}


\section{Introduction}

\begin{figure*}
\centering
\includegraphics[scale=0.55]{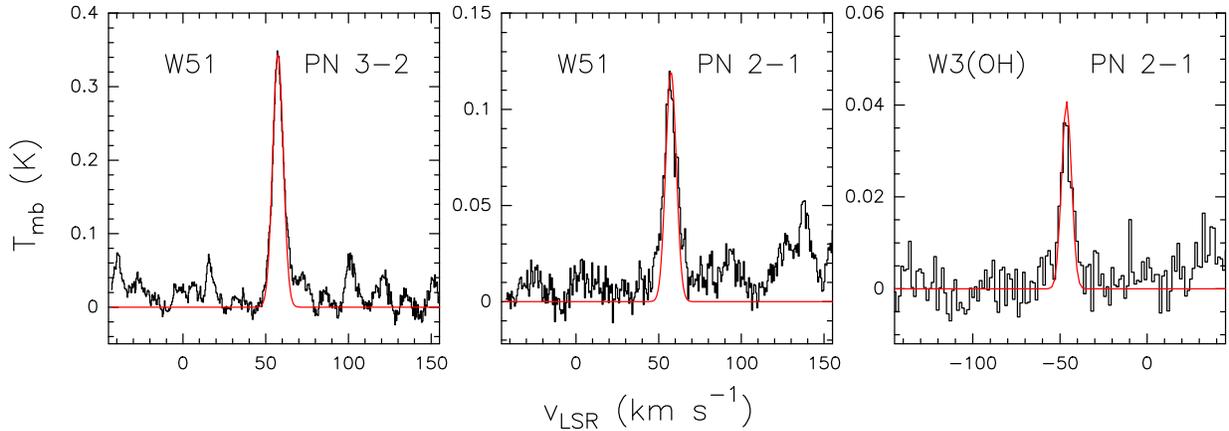}
\caption{PN transitions observed towards W51 and W3(OH). The red line is the LTE fit for a excitation temperature of 35 K (see text), and the source-average column densities given in Table \ref{table-physical-parameters}.}
\label{figure-PN}
\end{figure*}


\begin{table*}
\caption{Summary of the IRAM 30m observations towards the W51 and W3(OH) hot molecular cores.}
\begin{center}
\begin{tabular}{c c c c c}
\hline
 Date & Source & $RA_{\rm J2000}$ &  $DEC_{\rm J2000}$ & Frequency coverage \\
       & &   &   & (GHz) \\
\cline{2-3}
\hline       
2012 Apr 25 &  W51  & 19$h$ 23$m$ 43.90$s$ & 14$^{\circ}$ 30$^{\prime}$ 32.0$^{\prime\prime}$  & 88.83-96.60 $\&$  \\ \cline{2-4}
             &  W3(OH)  & 2$h$ 27$m$ 04.69$s$ & 61$^{\circ}$ 52$^{\prime}$ 25.51$^{\prime\prime}$  &  104.51-112.29 \\ 
         
\hline
2012 Dec 13   & W51  & 19$h$ 23$m$ 43.90$s$ & 14$^{\circ}$ 30$^{\prime}$ 34.8$^{\prime\prime}$ & 99.70-106.30 $\&$  \\ 
               &       &             &            & 238.20-245.98  \\ 
\hline
2015 Dec 9     & W51  & 19$h$ 23$m$ 43.90$s$ & 14$^{\circ}$ 30$^{\prime}$ 32.0$^{\prime\prime}$ & 134.47-142.25 $\&$ \\ 
                &      &             &            & 150.15-157.90  \\ 
\hline
\end{tabular}
\end{center}
\label{table-observations}
\end{table*}   


The detection of new interstellar molecules related with prebiotic chemistry in star-forming regions will allow us to make progress on understanding how the building blocks of life could originate in the interstellar medium (ISM). However, there is a key ingredient that still evades detection:
phosphorus, P. This element is essential for life (\citealt{macia97}), because it plays a central role in the formation of P-bearing macromolecules such as phospholipids, which are the structural components of all cellular membranes. Moreover, living cells on Earth use P-bearing compounds, phosphates, to transport energy in the form of adenotryphosphate (ATP; see e.g. \citealt{pasek05}). 
Especially important to basic biochemistry is the P$-$O bond, fundamental for many relevant biological molecules such as phosphate esters, which are essential for the formation of the backbone of the genetic macromolecule deoxyribonucleic acid (DNA).


Phosphorus is thought to be synthesized in massive stars and injected to the ISM through supernova explosions (\citealt{koo13}). It is a cosmically relatively abundant element, P/H$\sim$ 3$\times$10$^{-7}$ (\citealt{grevesse98}). 
It has been detected towards atmospheres of stars (\citealt{caffau11,roederer14,caffau16}), but it is barely detected in the ISM. The ion P$^+$ has been detected in several diffuse clouds (\citealt{jura78}), and only a few simple P-bearing species (PN, PO, CP, HCP, C$_2$P, PH$_3$) have been identified towards the circumstellar envelopes of very evolved objects (\citealt{tenenbaum07,debeck13,agundez14}). In star-forming regions, only PN has been detected so far (\citealt{turner87}; \citealt{ziurys87}; \citealt{fontani16}). Previous searches of PO towards star-forming regions (\citealt{sutton85}; \citealt{matthews87}) were unsuccesful.



\begin{figure*}
\centering
\includegraphics[scale=0.6]{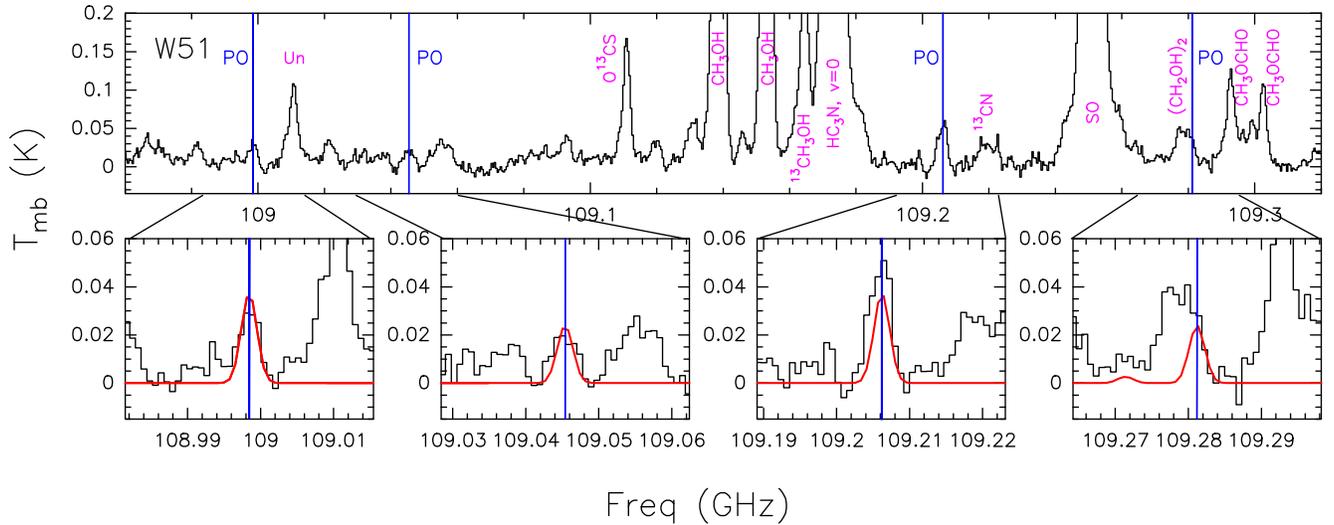}
\caption{Spectrum observed at 3 mm towards W51. The PO transitions are indicated with blue vertical lines. The lower panels show zoom-in views of the PO transitions. The red line is the LTE fit with a excitation temperature of 35 K and an assumed source size of 20$\arcsec$ (see text).}
\label{figure-PO-W51-3mm}
\end{figure*}

Despite the astrobiological importance of P-bearing molecules, their chemistry is still poorly understood in the ISM. 
The few theoretical models devoted to P-chemistry disagree in both the chemical formation pathways and the predictions of the abundances of PO and PN.
While some works (\citealt{millar87,adams90,charnley94}) suggest that PN should be more abundant than PO, other studies involving theoretical modeling and laboratory experiments 
(e.g. \citealt{thorne84}) predict that P should be found mainly in the form of PO.



To constrain the chemical models and understand the chemistry of phosphorus in the ISM, astronomical detections of PO and PN in star-forming regions are required. 
We present observations searching for PN and PO towards two massive star-forming regions: the W51 e1/e2 and the W3(OH) complexes (with luminosities of 1.5$\times$10$^{6}$ and 1.0$\times$10$^{5}$ L$_{\odot}$, respectively). 
The W51 star-forming region, located at a distance of 5.1 kpc (\citealt{xu09}) in the Sagittarius arm of our Galaxy, harbors two compact radio sources (\citealt{scott78}) known as e1 and e2, which are associated with hot molecular cores with a rich chemistry (\citealt{zhang97,zhang98,liu01,ikeda01,remijan02,demyk08,kalenskii10,lykke15}).
The W3(OH) complex, located at 2.04 kpc (\citealt{hachisuka06}), harbors two massive objects separated by 6$\arcsec$: the UC HII region W3(OH) (\citealt{wilner95}) and the chemically rich W3(H$_2$O) hot molecular core (\citealt{wyrowski99b}; \citealt{zapata11b}; \citealt{qin15}).

In this paper, we report the first detection of PO towards these two massive star-forming regions. 

\section{Observations}
\label{observations}

We used data from different observing runs performed with the IRAM 30m telescope at Pico Veleta (Spain) on 2012 and 2015\footnote{The Dec 2012 data were obtained from \citet{lykke15}.}. The dates, positions and frequency coverage of these observations are summarized in Table \ref{table-observations}. The W51 e1/e2 (hereafter W51) region was observed at 1, 2 and 3 mm, while the W3(OH)
region was observed at 3 mm.


All observing campaigns used the Eight Mixer Receiver (EMIR) and the Fast Fourier Transform Spectrometer (FTS), which provides a spectral resolution of 0.195 kHz.
The spectra were exported from the software package CLASS of GILDAS\footnote{The GILDAS software is available at http://www.iram.fr/IRAMFR/GILDAS} to MADCUBAIJ\footnote{Madrid Data Cube Analysis on ImageJ is a software developed in the Center of Astrobiology (Madrid, INTA-CSIC) to visualize and analyze single spectra and datacubes (Mart\'in et al., {\it in prep.}).}, which was used for the line identification and analysis. 
The line intensity of the IRAM 30m spectra was converted to the main beam temperature T$_{\rm mb}$ scale, using the 
efficiencies provided by IRAM\footnote{http://www.iram.es/IRAMES/mainWiki/Iram30mEfficiencies}. The half-power beam widths of the observations can be estimated using the expression $\theta_{\rm beam}$[arcsec]=2460/$\nu(\rm GHz)$.

\begin{figure}
\centering
\includegraphics[scale=0.40]{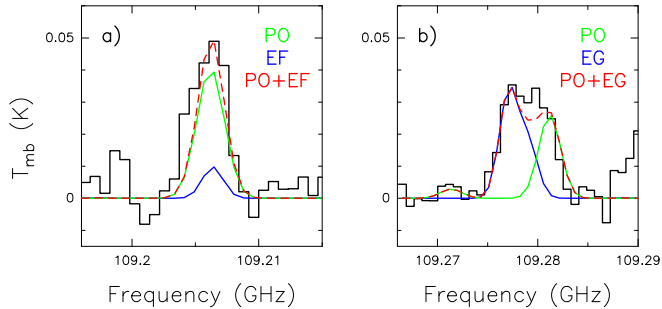}
\caption{Simultaneous LTE fit of the PO transitions at 109.206 GHz and 109.281 GHz and transitions of ethyl formate (EF) and ethylene glycol (EG) that are partially blended.}
\label{figure-PO-contamination}
\end{figure}


\section{Results}
\label{results}

\subsection{Detection of PN}


We have detected the $J$=2$-$1 and $J$=3$-$2 transitions of PN (Table \ref{table-lines-PN}) towards W51 (left and middle panels of Fig \ref{figure-PN}). This molecule was already detected towards this source by \citet{turner87} using the NRAO 12m telescope. 
The comparison of the line intensities of the PN $J$=2$-$1 transition observed with the 30m and 12m telescopes towards W51 allows us to estimate the size of the PN emission. The approximate size of the emitting region, $\theta_s$, can be estimated from the expression:

\begin{equation}
T_{\rm 12m}\frac{\theta_{\rm s}^2+\theta_{\rm 12m}^2}{\theta_{\rm s}^2}=T_{\rm 30m}\frac{\theta_{\rm s}^2+\theta_{\rm 30m}^2}{\theta_{\rm s}^2} ,
\end{equation}
where $\theta_{\rm 12m}$=69$\arcsec$ and $\theta_{\rm 30m}$=26$\arcsec$ are the HPBW of the two telescopes, $T_{\rm 12m}$=20 mK and $T_{\rm 30m}$=120 mK are the corresponding main-beam brightness temperatures.
We obtain an angular size of the emitting region of $\sim$12$\arcsec$. Using this angular diameter, the LTE simulated spectra reproduce the PN $J$=2$-$1 and $J$=3$-$2 transitions for an excitation temperature of $\sim$35 K (Fig. \ref{figure-PN}), as previously found by \citet{turner87}. We derived a PN source-averaged column density of 2.1$\times$10$^{13}$ cm$^{-2}$ (Table \ref{table-physical-parameters}). 
Using the derived $H_{\rm 2}$ column density $N_{\rm H_2}$ (see the details of the calculation in Appendix \ref{appendix}), the molecular abundance with respect to molecular hydrogen is 1.1$\times$10$^{-10}$.  

We have also detected the $J$=2$-$1 PN transition towards W3(OH) (right panel of Fig. \ref{figure-PN}). To our knowledge this is the first detection of PN towards this source. Assuming the same excitation temperature and linear diameter as that of W51 (i.e. 30$\arcsec$ at the distance of W3(OH)), we obtained a source-averaged PN column density of 0.2$\times$10$^{13}$ cm$^{-2}$, which implies a molecular abundance of 0.4$\times$10$^{-10}$.


\begin{table}
\caption{Spectroscopic parameters of targeted PN transitions, from CDMS. }
\vspace{3mm}
\begin{center}
\vspace{-4mm}
\begin{tabular}{c c c c  }
\hline
 Frequency (GHz) & Transition & S$_{\rm ij}\mu^{2}$ (D$^{2}$) & E$_{\rm up}$ (K)   \\      
\hline
93.97977 & 2$-$1 & 15.1 & 6.8  \\
140.96769 & 3$-$2 & 22.6 & 13.5  \\
\hline
\end{tabular}
\end{center}
\label{table-lines-PN}
\end{table}

\begin{table}
\tabcolsep 1.5pt
\caption{Physical parameters of PO and PN from the LTE analysis.}
\begin{center}
\begin{tabular}{c c c c c c c c}
\hline

Molecule & Source & $T_{\rm ex}$        &  $N_{\rm s}$   &  $N_{\rm b}$ &  $v_{\rm LSR}$  & $\Delta v$ &  $X_{\rm s}$ \\ 
         &  & (K)   &  \multicolumn{2}{c}{($\times$10$^{13}$ cm$^{-2}$)}       &  (km s$^{-1}$) & (km s$^{-1}$) & (10$^{-10}$)  \\
\hline\hline

PN   & W51  & 35     &  2.1     & 0.4 &  57.5    &  8.2 & 1.1  \\  
     & W3(OH)  & 35     &  0.2   & 0.13   &  -46.2    &  6.6 & 0.4  \\ 
\hline
PO    & W51  & 35   &  4.0  & 0.7     &  57.0    & 7.0    & 2.0 \\
     & W3(OH)  & 35     &  0.6  & 0.3     &  -46.2    &  6.6 & 1.2  \\ 
\hline
\end{tabular}
\end{center}
{\footnotesize{Note: $T_{\rm ex}$ (excitation temperature), $N_s$ (source averaged column density), $N_b$ (beam averaged column density), $v_{\rm LSR}$ (central velocity of the molecular emission), $\Delta v$ (molecular linewith, i.e. the full width half maximum of the line gaussian profile), $X_{\rm s}$ (source averaged molecular abundances).}}  \\
\label{table-physical-parameters}
\end{table}


\begin{table}
\caption{Spectroscopic parameters of targeted PO transitions (S$_{\rm ij}\mu^{2}$ $>$ 5 D$^{2}$ and  E$_{\rm up}$ (K) $<$ 300 K), from CDMS.}
\vspace{3mm}
\begin{center}
\vspace{-4mm}
\begin{tabular}{c c c c  }
\hline
 Frequency (GHz) & Transition & S$_{\rm ij}\mu^{2}$ (D$^{2}$) & E$_{\rm up}$ (K)   \\      
\hline
\multicolumn{4}{c}{J=5/2$-$3/2, $\Omega$=1/2} \\
\hline
108.99845 & F=3$-$2, l=e & 9.9 & 8.4  \\
109.04540 & F=2$-$1, l=e & 6.4 & 8.4  \\
109.20620 & F=3$-$2, l=f & 9.9 & 8.4  \\
109.28119 & F=2$-$1, l=f & 6.4 & 8.4  \\
\hline
\multicolumn{4}{c}{J=7/2$-$5/2, $\Omega$=1/2} \\
\hline
152.65698 & F=4$-$3, l=e & 13.6 & 15.7 \\
152.68028 & F=3$-$2, l=e & 10.1 & 15.7\\
152.85545 & F=4$-$3, l=f & 13.6 & 15.7\\
152.88813 & F=3$-$2, l=f & 10.1 & 15.7\\
\hline
\multicolumn{4}{c}{J=11/2$-$9/2, $\Omega$=1/2} \\
\hline
239.94898 & F=6$-$5, l=e & 20.9 & 36.7 \\
239.95810 & F=5$-$4, l=e & 17.4 & 36.7 \\
240.14106 & F=6$-$5, l=f & 20.9 & 36.7 \\
240.15253 & F=5$-$4, l=f & 17.4 & 36.7 \\
\hline
\end{tabular}
\end{center}
\label{table-lines-PO}
\end{table}

\begin{figure*}
\centering
\includegraphics[scale=0.55]{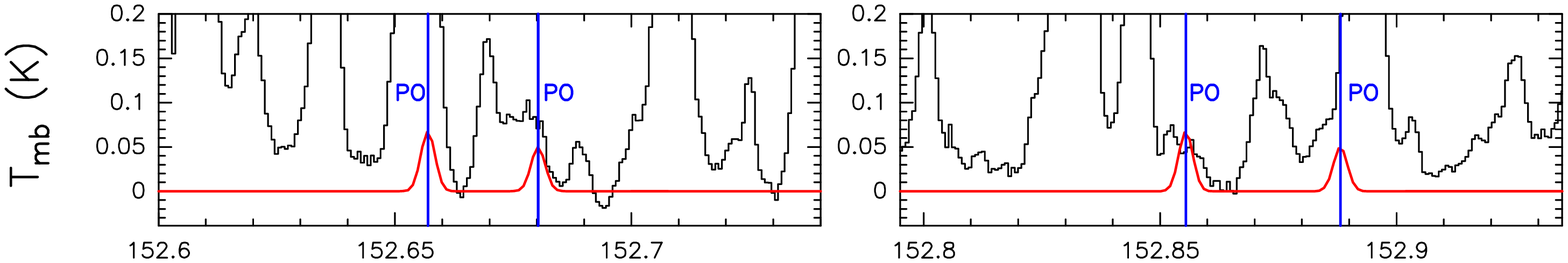}
\vskip5mm
\includegraphics[scale=0.55]{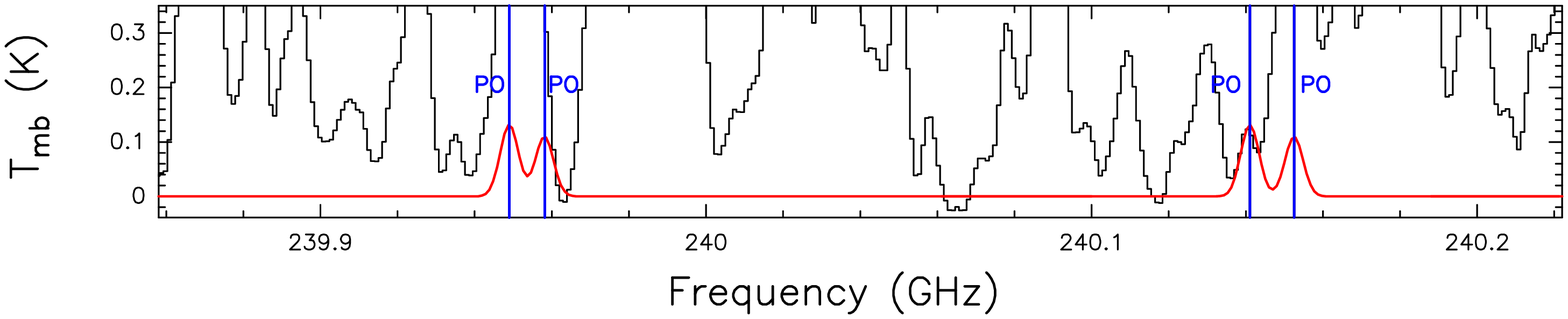}
\caption{PO transitions at 1 and 2 mm observed towards the W51 hot core. The PO transitions are indicated with blue vertical lines and the red line is the LTE fit as in Fig. \ref{figure-PO-W51-3mm}.}
\label{figure-PO-W51}
\end{figure*}

\subsection{Detection of PO}

The PO radical is a diatomic molecule with a dipole moment of $\mu$=1.88 D (\citealt{kanata88}). Its millimeter spectrum was measured in the gas phase in the laboratory by \citet{kawaguchi83}. 
PO has a $^2\Pi_{\rm r}$ ground electronic state, and therefore each $J$-transition splits into a doublet because of $\Lambda$ coupling. The spin of I$=$1/2 of the phosphorus atom produces a further splitting of each $\Lambda$ doublet into two hyperfine levels.
Table \ref{table-lines-PO} shows the PO transitions that are most likely detectable in the ISM with the 30m telescope: the brightest transitions of the $\Omega$=1/2 ladder, with S$_{\rm ij}\mu^{2}$ $>$ 5 D$^{2}$, and with energy levels  E$_{\rm up}$ $<$ 300 K.
The lower energy (E$_{\rm up}$= 3 K) and lower frequency transitions ($J$=3/2$-$1/2) at 65 GHz of the $\Omega$=1/2 ladder are not observable due to atmospheric opacity. 
The $J$=9/2$-$7/2 quadruplet falls at 196 GHz, out of the range covered by the IRAM 30m telescope. 
The search for higher energy PO transitions (e.g. J=13/2$-$11/2 at 283 GHz) is much less feasible, due to strong contamination from brighter molecular lines at this frequency. 
Therefore, the PO transitions given in Table \ref{table-lines-PO} are the best candidates for a detection. The observations towards W51 covered all these PO transitions (at 1, 2 and 3 mm), while the observations towards W3(OH) only covered the 3 mm transitions.



We have used MADCUBAIJ to identify the PO transitions in the observed spectra. We have assumed the same temperature ($T_{\rm ex}$=35 K) and source size as that of PN to produce the LTE spectrum.
We show the 3 mm spectra of W51 and the LTE fit in Fig. \ref{figure-PO-W51-3mm}. We have clearly identified the PO quadruplet towards W51. The four transitions are well fitted by the LTE model. The lines are detected above 6$\sigma$. We note that these PO transitions have already been detected in the wind of the oxygen-rich AGB star IK Tauri by \citet{debeck13}. 

The PO transitions at 108.99845 GHz and 109.04540 GHz are completely free of contamination. The transition at 109.20620 GHz also seems unblended, although the LTE model slightly underestimates the observed intensities. The transition at 109.28119 GHz appears clearly blended with another species. To firmly confirm the detection of PO, we have evaluated the possibility of contamination from lines of other molecules. We have searched in the available catalogs (CDMS\footnote{http://www.astro.uni-koeln.de/cdms} and JPL\footnote{http://spec.jpl.nasa.gov/}) and conclude that there are no transitions blended with the PO lines at 108.99845 GHz and 109.04540 GHz. The only possible contaminantion of the PO transition at 109.20620 GHz is the complex organic molecule ethyl formate, C$_2$H$_5$OCHO.
Interestingly, this molecule has been recently identified towards W51 (Rivilla et al., {\it submitted}) based on the detection of multiple transitions, which has allowed us to derive its rotational temperature and column density. Using the physical parameters derived in Rivilla et al. ({\it submitted}) ($N$=10$^{16.3}$, $T_{\rm}=$100 K), we have predicted the LTE line profile of the ethyl formate transition close to the 109.20620 GHz transition of PO. The result is shown in the left panel of Fig. \ref{figure-PO-contamination}. Ethyl formate has a very weak contribution to the observed emission, which is hence mainly due to PO. The LTE profile reproduces very well the observed line profile if the emission of both molecules is taken into account (see left panel of Fig. \ref{figure-PO-contamination}). 

In the case of the 109.28119 GHz transition of PO, we have found that it is blended with a line of ethylene glycol, (CH$_2$OH)$_2$. This species has been tentatively identified towards W51 by \citet{lykke15} from observations at 1 mm. Our 2 and 3 mm spectra clearly confirm the detection of ethylene glycol, and allow us to confirm the excitation temperature of $\sim$120 K previously obtained by \citet{lykke15}. We show in the right panel of Fig. \ref{figure-PO-contamination} the LTE model of the ethylene glycol transition close to the 109.28119 GHz transition of PO. The combined contribution of PO and ethylene glycol reproduces well the observed spectral line profile.

The LTE fit of the four PO spectral features at 3 mm gives a source-averaged PO column density of 4$\times$10$^{13}$ cm$^{-2}$ and a molecular abundance of 2$\times$10$^{-10}$ (Table \ref{table-physical-parameters}). In Fig. \ref{figure-PO-W51} we also show the predicted LTE lines profile of the PO at 1 and 2 mm (Table \ref{table-lines-PO}). The LTE prediction is consistent with the observed spectra. The predicted PO transition at 240.14106 GHz fits a partially blended line of the observed spectrum (Fig. \ref{figure-PO-W51}), while the other PO transitions appear heavily blended with brighter lines of other molecular species.  
In summary, we have detected all the 12 detectable PO transitions towards W51 with intensities consistent with LTE emission: two of them (at 3 mm) are unblended, two (at 3 mm) are partially blended, while the others (all the 2 mm and 1 mm transitions) are heavily blended. 

\begin{figure*}
\centering
\includegraphics[scale=0.6]{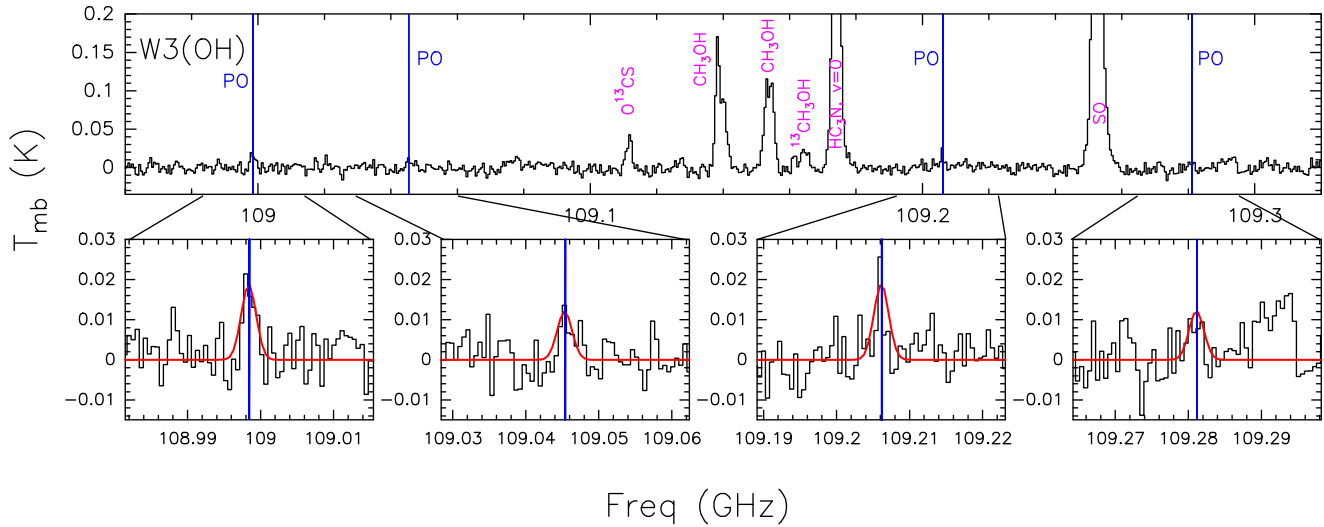}
\caption{Same as Fig. \ref{figure-PO-W51-3mm} for the W3(OH) hot core.}
\label{figure-PO-W3OH}
\end{figure*}
\begin{figure}
\centering
\includegraphics[scale=0.3]{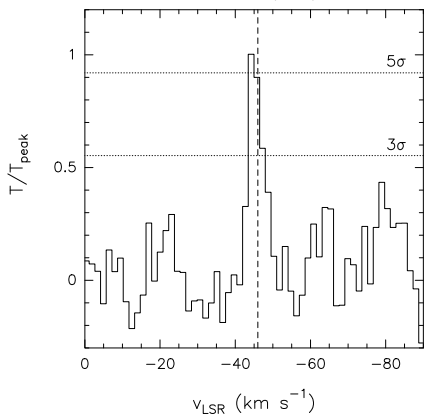}
\caption{Composite normalized spectrum of PO towards W3(OH). The dashed vertical line indicates the systemic velocity of the source, while the dotted horizontal lines show the 3$\sigma$ and 5$\sigma$ detection levels.}
\label{figure-PO-W3OH-composite}
\end{figure}

For W3(OH), we show in Fig. \ref{figure-PO-W3OH} the observed 3 mm spectra. The two brightest lines of PO (108.99845 GHz and 109.20620 GHz) are detected at a $\sim$ 4$\sigma$ level. We have fitted these lines with the LTE model, assuming $T_{\rm ex}$=35 K and the same diameter assumed for PN (30$\arcsec$), and obtained a column density of 0.6$\times$10$^{13}$ cm$^{-2}$ and a molecular abundance of 1.2$\times$10$^{-10}$ (Table \ref{table-physical-parameters}). 
To confirm this detection we have stacked the 3 mm observed spectra to improve the signal to noise ratio.
This method is used to identify molecular species marginally detected in several lines. In practice, we have averaged the observed spectra of the 4 PO transitions with the weights calculated assuming LTE, and normalized the result to the peak value. The resulting spectrum, shown in Fig. \ref{figure-PO-W3OH-composite}, indicates a clear detection at the systemic velocity of the source. The signal is $>$ 5$\sigma$, confirming the detection of PO towards W3(OH).

\section{Discussion: the chemistry of phosphorus}

\subsection{The PO/PN ratio}


The LTE fits of PO and PN transitions show that the molecular abundances of the two molecules are similar in the two sources ($\sim$10$^{-10}$), being the PO abundance higher by a factor $\sim$2. 
Our detections towards W51 and W3(OH) show that the line intensities of PO at 3 mm are a factor $\geq$4 weaker than the PN $J$=2$-$1 transition at similar wavelengths. This makes the identification of PO more difficult than that of PN and explains why previous attempts to detect PO were fruitless.
\citet{matthews87} derived upper limits of the beam averaged column densities\footnote{\citet{matthews87} derived the PO upper limits assuming a dipole moment of PO of 2.13 Debye. \citet{kanata88} measured a dipole moment of 1.88, so we have corrected the PO upper limits of Table 1 of \citet{matthews87} by multiplying for a factor (2.13/1.88)$^{2}\simeq$1.3.} of four hot molecular cores of $<$ (0.2$-$2.7)$\times$10$^{13}$ cm$^{-2}$. These upper limits are consistent with the PO column densities we have derived, suggesting that the non-detections of \citet{matthews87} were due to insufficient sensitivity. 

More recently, \citet{fontani16} have found upper limits for the PO/PN ratio of $<$ (1.3$-$4.5) towards a sample of massive star-forming cores (see Table \ref{table-PO-PN-ratio}), which are also consistent with the abundances of both P-bearing species estimated by us. The upper limit of the PO/PN ratio of $\sim$0.5$-$1.3 found towards a protostellar shock by \citet{aota12} is also in agreement with our observed ratios.

Before the present detections, PO had only been detected towards the envelopes of two evolved stars, VY Canis Majoris (\citealt{tenenbaum07}), and IK Tauri (\citealt{debeck13}), where the PO/PN ratios are 2.2 and 0.17$-$2, respectively. These ratios are similar to those found towards W51 and W3(OH), suggesting that phosphorus seems to be equally distributed in the form of PO and PN in both interstellar and circumstellar material.




\begin{table}
\tabcolsep 1.0pt
\caption{Column densities and relative abundances of PO and PN in different sources.}
\begin{center}
\begin{tabular}{c c c c c }

\hline
Source &  $N_{\rm PN}$         &  $N_{\rm PO}$  & $N_{\rm PO}$/$N_{\rm PN}$ &  Ref. \\
\hline
\multicolumn{5}{c}{Star-forming regions}  \\
\cline{2-3}
 &  $\times$10$^{11}$ cm$^{-2}$         &  $\times$10$^{11}$ cm$^{-2}$  &  &  \\
\hline
W51 e1/e2      &  210   &   400     &    1.8   &  (1)\\
W3(OH)       &   20     &   60       &   3.0    &  (1)   \\
AFGL5142-EC$^{a}$      &  8.3    & $<$ 20      &   $<$ 2.4 &   (2)      \\   
05358$-$mm3$^{a}$      &  2.9   &   $<$ 13     &  $<$ 4.5   &  (2)   \\ 
AFGL5142$-$MM$^{a}$      &  6.9   & $<$ 12      & $<$ 1.7   &  (2)    \\ 
18089$-$1732$^{a}$      &  3.3   &   $<$ 6.3    & $<$ 1.9   &  (2)    \\ 
18517$-$0437$^{a}$      &  2.0   &  $<$ 4.4    &  $<$ 2.2   & (2)   \\ 
G5.89-0.39$^{a}$      &   10.4  &  $<$ 14      &  $<$ 1.3   &  (2)   \\ 
19410+2336$^{a}$      &   1.6  &   $<$ 3.3      & $<$ 2.0   & (2)    \\ 
 ON1$^{a}$     &    3.8     & $<$  6.3    &  $<$ 1.7 & (2) \\   
\hline
\multicolumn{5}{c}{Protostellar shocks}  \\
\hline
L1157 B1$^{a}$      &  4.2$-$11   &  $<$ 5.3     & $<$ 0.5$-$1.3   &  (3)   \\ 
\hline    
\multicolumn{5}{c}{Envelopes of evolved stars}  \\
\cline{2-3}
 &  $\times$10$^{15}$ cm$^{-2}$         &  $\times$10$^{15}$ cm$^{-2}$  &  &  \\
\hline
VY Canis Majoris      &  1.2   &   2.8     &   2.2    &  (4)  \\ 
IK Tauri      &     2.2     &  0.4$-$4.4  & 0.17$-$2   &  (5) \\ 

\hline
\end{tabular}
\end{center}
\vspace{-3mm}
{$^{a}$ Beam-averaged column densities.} \\
{{\it References:} (1) This paper; (2) \citet{fontani16}; (3) \citet{aota12}; (4) \citet{tenenbaum07}; (5) \citet{debeck13}.} \\

\label{table-PO-PN-ratio}
\end{table}

\subsection{Chemical model: constraining the phosphorus chemistry}


We have investigated the formation of the phosphorus-bearing molecules PO and PN in massive star-forming regions using the model based on \citet{vasyunin13}, which simulates the chemical evolution of a parcel of gas and dust with time-dependent physical conditions. First, the formation of a dark dense clump from a translucent cloud during free-fall collapse is simulated ($''$cold starless phase $''$). During this phase, the density increases from 10$^3$ cm$^{-3}$ to 10$^{6}$ cm$^{-3}$ and visual extinction rises from A$_v$=2 to A$_v\gg$100 mag. Gas and dust temperatures are both decreasing slightly from 20 K to 10 K. 
 On a second phase, the dense dark clump warms up from 10 K to 200 K during 2$\times$10$^{5}$ years, thus developing into a hot core ($''$warm-up protostellar phase $''$). We have defined as time 0 the moment when the protostar starts to heat up the environment.

We used two sets of initial chemical abundances to explore the influence of the poorly constrained P depletion factor: i) the $''$low metals $''$ model (EA1, see Table 1 in \citealt{wakelam08}), where the atomic P is depleted and has an abundance of 2$\times$10$^{-10}$ with respect to hydrogen, typically used in dark cloud chemical models; and ii) the $''$high metals $''$ model, where the phosphorus abundance is increased by a factor of $\sim$25 to 5$\times$10$^{-9}$.

\begin{figure*}
\centering
\includegraphics[scale=0.60, angle=0]{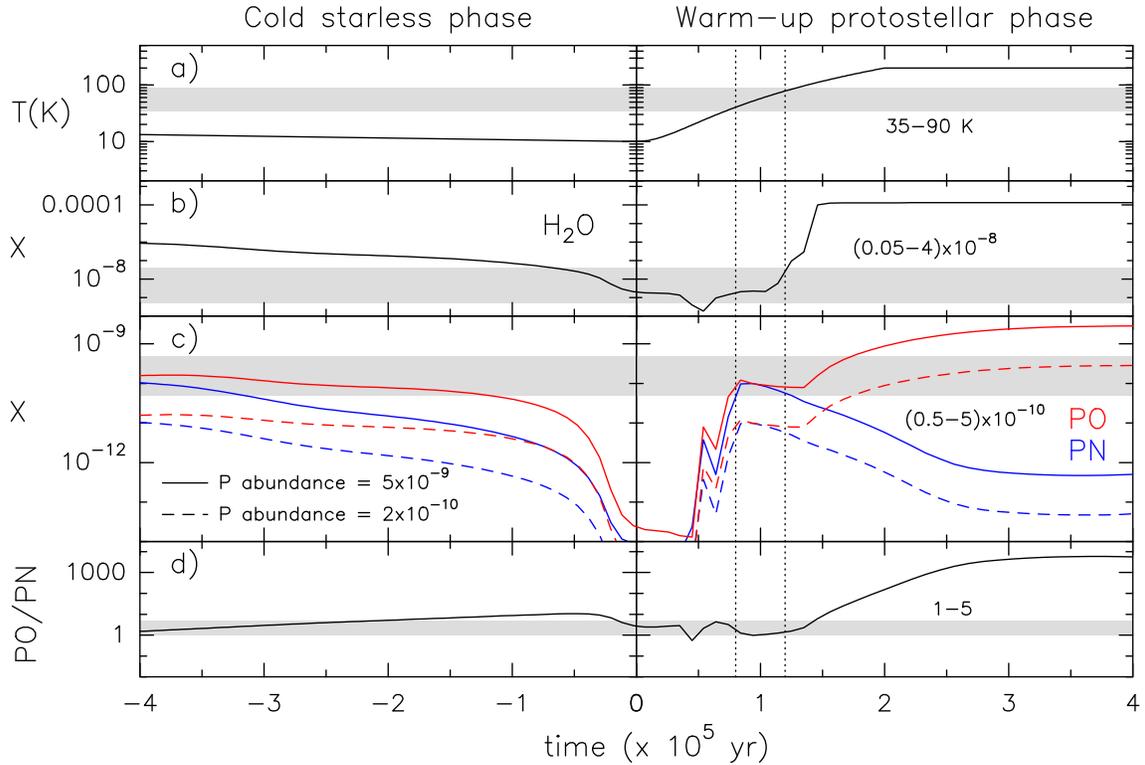}
\caption{\small Results of our chemical model. The left panels correspond to the cold collapse phase, and the right panels to the subsequent warm-up phase. We show the evolution during 0.8$\times$10$^{5}$ yr, being time 0 the moment when the protostar starts to heat up the environment. The two vertical dotted lines in the right panels indicate the temporal range for which the observational constraints are fulfilled (see text).
 {\it a)} Evolution of the temperature assumed by the model. The grey band corresponds to the temperature range between 35 and 90 K which is compatible with the observational constraints (see text).  
{\it b)} H$_2$O abundance. The grey band corresponds to the range found towards massive star-forming regions by \citet{marseille10}.
{\it c)} PO and PN abundances predicted by the model (red and blue lines, respectively). We have considered two different initial abundances of depleted atomic phosphorus: 5$\times$10$^{-9}$ (solid lines, "high metals" model), and 2$\times$10$^{-10}$ (dashed lines, "low metals" model, usually assumed in dark cloud models). 
The grey band indicates a conservative range of one order of magnitude around the molecular abundances derived from our observations of PO and PN (see text).
{\it d)} Ratio of the molecular abundances of PO and PN. The grey band shows a conservative range (1$-$5) around the ratio found by the observations.}
\label{figure-model}
\end{figure*}

The results of the modeling are shown in Fig. \ref{figure-model}, which shows the temporal evolution of the temperature (Fig. \ref{figure-model}a), gas abundance of water (Fig. \ref{figure-model}b), gas abundances of PO and PN (Fig. \ref{figure-model}c) and the PO/PN ratio (Fig. \ref{figure-model}d).
The two P-bearing species are chemically related and are formed purely in a sequence of gas-phase ion-molecule and neutral-neutral reactions. PO is efficiently formed during the cold collapse phase in a sequence of reactions:

\begin{equation}
{\rm H_3O^+ + P  \longrightarrow HPO^+ + H_2}
\end{equation}
\begin{equation}
{\rm HPO^+ + e^- \longrightarrow PH + O}
\end{equation}
\begin{equation}
{\rm HPO^+ + e^- \longrightarrow PO + H}
\end{equation}
\begin{equation}
{\rm P^+ + H_2  \longrightarrow PH_2^+}
\end{equation}
\begin{equation}
{\rm PH_2^+ + e^- \longrightarrow PH + H}
\end{equation}
\begin{equation}
{\rm O + PH \longrightarrow PO + H}
\end{equation}

PN is mainly formed from PO via:

\begin{equation}
{\rm N + PO \longrightarrow PN + O}
\end{equation}

During the starless phase the model predicts that PO is more abundant than PN (Fig. \ref{figure-model}c). Both species freeze out to grains at the end of the cold collapse phase and consequently the gas abundances sharply decrease (Fig. \ref{figure-model}c). PO has an abundance $\sim$5 times higher than PN at that time. 


At the warm-up phase, once the temperature gradually increases from 10 K to 200 K (Fig. \ref{figure-model}a), the species evaporate to the gas phase in the order determined by their desorption energies. PO and PN desorb simultaneously when the temperature reaches $\sim$35 K ($\sim$0.7$\times$10$^{5}$ yr, Fig. \ref{figure-model}c), because in our model they have the same evaporation energy of 1900 K\footnote{To the best of our knowledge, there are no experimental data on PN and PO desorption energies, and therefore, the results of the modeling should be accepted with some caution.}.

Some PO is converted to PN via gas-phase reactions, and the abundances of the two species become almost equal until $\sim$1.2$\times$10$^{5}$ yr (Fig. \ref{figure-model}c). Once the temperature reaches $\sim$100 K, water ice evaporates and the gas-phase abundance of water reaches a high value of 10$^{-4}$ (see Fig. \ref{figure-model}b). Then, the abundance of protonated water H$_3$O$^+$ correspondingly increases, and PN reacts with H$_3$O$^+$:

\begin{equation}
{\rm PN + H_3O^+ \longrightarrow HPN^+ + H_2O}
\end{equation}

In turn, HPN$^+$ has two equally probable channels of dissociative recombination:

\begin{equation}
{\rm HPN^+ + e^- \longrightarrow PN + H}
\end{equation}
\begin{equation}
{\rm HPN^+ + e^- \longrightarrow PH + N}
\end{equation}

Since the abundance of atomic oxygen is higher by almost an order of magnitude than the abundance of atomic nitrogen, PH is preferrably converted to PO rather than PN. As such, PN is gradually destroyed, while PO is additionally produced, which significantly increases the PO/PN ratio (Fig. \ref{figure-model}d).

To compare the results of the observations with the predictions of our chemical models, we show the values of the abundances and the PO/PN ratio derived from the observations (Table \ref{table-physical-parameters}) in Figs.~\ref{figure-model}c and \ref{figure-model}d, respectively. 
The main source of uncertainty of the estimated abundances arises from the uncertainty of the size of the emmiting region and the H$_2$ column density. We have considered a conservative range of one order of magnitude uncertainty for the abundances, i.e. (0.5$-$5)$\times$10$^{-10}$ (see Fig.~\ref{figure-model}c). In the case of the PO/PN ratio, which is independent of the H$_2$ column density, we have considered a lower uncertainty of a factor of 5, i.e., values in the range 1$-$5. These observational constraints can only be explained by the "high metals" model (Fig. \ref{figure-model}c), and during a period of $\sim$5$\times$10$^{4}$ yr (between 0.8 and 1.2$\times$10$^{5}$ yr in Fig. \ref{figure-model}). This suggests that the value of the initial depleted atomic abundance of P is a factor of $\sim$25 higher (i.e. 5$\times$10$^{-9}$) than that commonly assumed by the standard "low metal" model. 

The temporal window that fulfills the observational constraints translates in a temperature range between the desorption temperature of the P-bearing molecules (35 K) and $\sim$90 K (see Fig.~\ref{figure-model}a). The former temperature is the same than that derived from PN transitions towards W51, which may indicate that the observations have detected the bulk of PN (and PO) that have recently desorbed from dust grains. The latter temperature is just below the sublimation temperature of the bulk of water (100 K), when PN is efficiently destroyed and partially converted to PO. Then, the gas abundace of water predicted by the model at this stage is still low, $\sim$10$^{-9}$, which is in good agreement with the values of 5$\times$10$^{-10}$ $-$ 4$\times$10$^{-8}$ found towards several massive star-forming regions by \citet{marseille10} (see Fig.~\ref{figure-model}b).  

\label{discussion}

\section{Summary and conclusions}

We report on the first detection of the key prebiotic molecule PO towards two star-forming regions: W51 e1/e2 and W3(OH). The derived molecular abundances of PO are $\sim$10$^{-10}$ in both sources. We have found an abundance ratio PO/PN of 1.8 and 3 for W51 e1/e2 and W3(OH), respectively, in agreement with the values estimated for evolved stars. 
Our chemical modelling indicates that the two molecules are chemically related and are formed via gas-phase ion-molecular and neutral-neutral reactions during the cold collapse. The molecules freeze out onto grains at the end of the collapse, and evaporate during the warm-up phase once the temperature reach $\sim$35 K.
Similar abundances of PO and PN are expected during a period of $\sim$5$\times$10$^4$ yr at the early stages of the warm-up phase, when the temperature is in the range 35$-$90 K. The observed molecular abundances require a relatively high initial abundance of depleted atomic phosphorus of 5$\times$10$^{-9}$, 25 times higher than the $"$low metal$"$ P-abundance typically used in dark cloud chemical models.   


\vspace{1cm}

\section*{Acknowledgments}
\begin{small}
The authors are grateful to the IRAM staff for its help during the observations of the IRAM 30m data, and especially to the Director for allowing us to use the telescope time during the DDT proposal D04$-$15.
This work was partly supported by the Italian Ministero dell'Istruzione, Universit\`a e Ricerca through the grant Progetti Premiali 2012 $-$ iALMA. PC and AV acknowledge the financial support of the European Research Council (ERC; project PALs 320620). JM-P acknowleges partial support by the MINECO under grants AYA2010-2169-C04-01, FIS2012-39162-C06-01, ESP2013-47809-C03-01 and ESP2015-65597-C4-1.
\end{small}


\bibliographystyle{mn2e}
\bibliography{biblio}

\begin{appendices}

\appendix{}
\section{Calculation of the hydrogen column density}
\label{appendix}

One can derive the hydrogen column density of a clump from the continuum flux emitted by the dust. Assuming an homogeneous clump with temperature $T_{\rm d}$ and opacity $\tau_{\rm \nu}$ at frequency $\nu$, the solution of the radiative transfer equation gives the intensity emitted:

\begin{equation}
I_{\rm \nu}=B_{\rm \nu}(T_{\rm d})(1-e^{-\tau_{\rm \nu}}),
\label{eq-intensity}
\end{equation}
where $B_{\rm \nu}$ is the Plank function:

\begin{equation}
B_{\rm \nu}=\frac{2h\nu^3}{c^2}\frac{1}{e^{h \nu/k T_{\rm d}}-1},
\end{equation}
where $c$ is the speed of light, $k$ is the Boltzmann's contant, and $h$ is the Plank's constant.

One can relate the flux density measured within the beam solid angle of the telescope ($\Omega$) with the intensity using:

\begin{equation}
I_{\rm \nu}=\frac{F_{\rm \nu}}{\Omega}   .
\label{eq-intensity-beam}
\end{equation} 

Combining eqs. \ref{eq-intensity} and \ref{eq-intensity-beam}, one can derive the opacity of the dust emission inside the beam:

\begin{equation}
\tau_{\rm \nu}= -{\rm ln} \left(1-\frac{F_\nu}{\Omega \hspace{1mm} B_\nu(T_d)}\right)   .
\label{eq-opacity}
\end{equation} 

Since the opacity is defined as:

\begin{equation}
\tau_{\rm \nu}=\int \kappa_\nu \hspace{1mm}  \rho \hspace{1mm}  dr,
\end{equation} 
where $\rho$ is the volume density and $\kappa_\nu$ is the absorption coefficient per unit density, one can derive the hydrogen column density with the optical depth using:

\begin{equation}
N_{H_2}=\int{\frac{\rho}{\mu \hspace{1mm} m_{\rm H}}\hspace{1mm}  dr} =\frac{\tau_\nu}{\mu \hspace{1mm}  m_{\rm H} \hspace{1mm} \kappa_\nu},
\label{eq-column-density}
\end{equation}
where $m_{\rm H}$ is the hydrogen mass and $\mu$=2.8 is the mean molecular mass of the ISM with respect to H$_{\rm 2}$ molecules. We have used eqs. \ref{eq-opacity} and \ref{eq-column-density} to derive the dust opacity and the hydrogen colum density of W51 and W3(OH) from the dust continuum emission from ATLASGAL (870 $\mu$m, \citealt{csengeri14}) and SCUBA (850 $\mu$m, \citealt{difrancesco08}) surveys, respectively. We have considered that the dust temperature is the same as the excitation temperature derived for PN, 35 K. We have adopted a gas-to-dust ratio of 100 and $\kappa_\nu$=1.85 cm$^{2}$ g$^{-1}$, from interpolation of the column 5 of Table 1 of \citet{ossenkopf94}. We have used the peak flux densities shown in Table \ref{table-column-densities}, and the corresponding telescope beams: 19.2$\arcsec$ (ATLASGAL) and 22.9$\arcsec$ (SCUBA). The results are given in Table \ref{table-column-densities}. We have used the values of $N_{\rm H_2}$ to compute the molecular abundances of PN and PO (see Table \ref{table-physical-parameters}).

\begin{table}
\tabcolsep 2.2pt
\caption{Continuum fluxes of W51 e1/e2 and W3(OH) clumps from ATLASGAL and SCUBA, respectively, and derived dust opacities and molecular hydrogen column densities.}
\begin{center}
\begin{tabular}{c c c c c }
\hline

Source & $\lambda$ ($\mu$m) & $F_{\rm \nu}$ (Jy beam$^{-1}$)       &  $\tau$   &  $N_{\rm H_2}$ ($\times$10$^{23}$ cm$^{-2}$)  \\ 
\hline\hline
W51     & 870 & 70.2$^{a}$     &  0.02     & 2.0   \\  
W3(OH)  & 850 & 26.3$^{b}$     &  0.005   & 0.5   \\ 
\hline
\end{tabular}
\end{center}
{\footnotesize{$^{a}$ From ATLASGAL (\citealt{csengeri14}); $^{b}$ From SCUBA (\citealt{difrancesco08})}.}  \\
\label{table-column-densities}
\end{table}

\end{appendices}

\bibliographystyle{mn2e}
\bibliography{bib}


\end{document}